\documentclass[preprint,pra,superscriptaddress]{revtex4-1}

\usepackage[dvips]{graphicx} 
\usepackage{amsfonts}
\usepackage{amssymb}
\usepackage{amscd}
\usepackage{amsmath}    
\usepackage{enumerate}
\usepackage{epsfig}
\usepackage{subfigure}
\usepackage{xcolor}

\begin{document}

\title{Effect of imperfect Faraday mirrors on security of a Faraday-Michelson quantum cryptography system}

\author{Wei-Long Wang}
\affiliation{State Key Laboratory of Mathematical Engineering and Advanced Computing, 450001, Zhengzhou, Henan, China}
\author{Ming Gao}
\email{Electronic address: gaoming.zhengzhou@gmail.com}
\affiliation{State Key Laboratory of Mathematical Engineering and Advanced Computing, 450001, Zhengzhou, Henan,
China}
\author{Zhi Ma}
\affiliation{State Key Laboratory of Mathematical Engineering and Advanced Computing, 450001, Zhengzhou, Henan,
China}

\begin{abstract}
The one-way Faraday-Michelson system is a very useful practical
quantum cryptography system where Faraday mirrors(FMs) play an
important role. In this paper we analyze the security of this system
against imperfect FMs. We consider the security loophole caused by
the imperfect FMs in Alice's and Bob's security zones. Then we
implement a passive Faraday mirror attack in this system. By
changing the values of the imperfection parameters of Alice's FMs,
we calculate the quantum bit error rate between Alice and Bob
induced by Eve and the probability that Eve obtains outcomes
successfully. It is shown that the imperfection of one of Alice's
two FMs makes the system sensitive to the attack. At last we give a
modified key rate as a function of the Faraday mirror imperfections.
The security analysis indicates that both Alice's and Bob's
imperfect FMs can compromise the secure key.

\end{abstract}

\maketitle
\section{INTRODUCTION}
Quantum key distribution (QKD) \cite{1,2} is one of the most
realistic applications in quantum information. It can generate
secure keys between two distant parties, commonly known as Alice and
Bob. The unconditional security has been proven even when an
eavesdropper, Eve, has unlimited computation power permitted by
quantum mechanics \cite{3,4,5,6}. However,  in practical QKD systems
there are always some imperfections that will leave loopholes for
Eve to take use of. Therefore, various hacking attacks based on the
imperfections of practical QKD systems are proposed \cite{7}. There
are Trojan horse attack \cite{8}, fake state attack \cite{9},
phase-remapping attack \cite{10}, time-shifted attack \cite{11} and
blinding attack\cite {12}. Passive Faraday-mirror attack based on
the imperfection of Faraday-mirrors in the two-way systems was
proposed in 2011 \cite{13}.

Since in two-way system, such as plug-and-play system \cite{14},
Alice admits pulses from other zones in, it is vulnerable to a
Trojan horse attack by Eve. Thus one-way QKD system has an obvious
advantage in security over two-way system. In 2005, a novel one-way
Faraday-Michelson quantum cryptography (FMQC) system \cite{15} was
proposed. The simple diagram of FMQC system with Eve is shown in
Fig. 1. It is an intrinsically stable QKD system free of fiber
birefringence, which makes it an important and useful practical QKD
system. In Alice's zone, a laser pulse is split into two pulses by
coupler $C_A$. One is transmitted on the short arm (denoted as time
mode $a$), and the other is transmitted on the long arm (denoted as
time mode $b$). There is a phase modulator $PM_A$ which modulates
the phase of the pulse transmitted on the long arm. The pulses are
coupled by coupler $C_A$ and transmitted to Bob. After arriving at
coupler $C_B$, the pulses are split into two
 groups and reflected back by two Faraday mirrors, respectively. The group transmitted on the long arm is
 modulated by phase modulator $PM_B$.

In this system four Faraday mirrors are used to compensate for any
birefringence effect in fibers automatically and perfectly. A
perfect FM is a combination of a $45^\circ$ Faraday rotator and a
reflecting mirror whose Jones matrix is given by $FM= \left(
\begin{array}{cc}
 0 & -1  \\
 -1 & 0\end{array} \right)$.
Thus after a state goes through a FM, the polarization becomes
orthogonal to that of the incoming state. The FM can automatically
compensate for any birefringence effect in fibers. For example, when
a photon passes through a birefringent medium forward and is
reflected by a FM backward. The matrices can be written as
\begin{equation}
T(\mp\theta)=\left(
\begin{array}{cc}
 cos(\theta) & \pm sin(\theta)  \\
 \mp sin(\theta) & cos(\theta)\end{array} \right)\left(
\begin{array}{cc}
 e^{i\varphi_o} & 0  \\
 0 & e^{i\varphi_e}\end{array} \right)\left(
\begin{array}{cc}
 cos(\theta) & \mp sin(\theta)  \\
 \pm sin(\theta) & cos(\theta)\end{array} \right),
 \end{equation}
 where $\varphi_o$ and $\varphi_e$ are the propagation phases of ordinary and
extraordinary rays, and $\theta$ is the rotation angle between the reference basis and the eigenmode basis of the
birefringent medium. The overall Jones matrix for a round trip is given by
\begin{equation}
T(-\theta)\times FM\times T(\theta)=e^{i(\varphi_o+\varphi_e)}FM,
\end{equation} which shows clearly that a perfect FM can compensate for any birefringence effect in fibers.

 \begin{figure}[!t]
 \includegraphics*[scale=0.88]{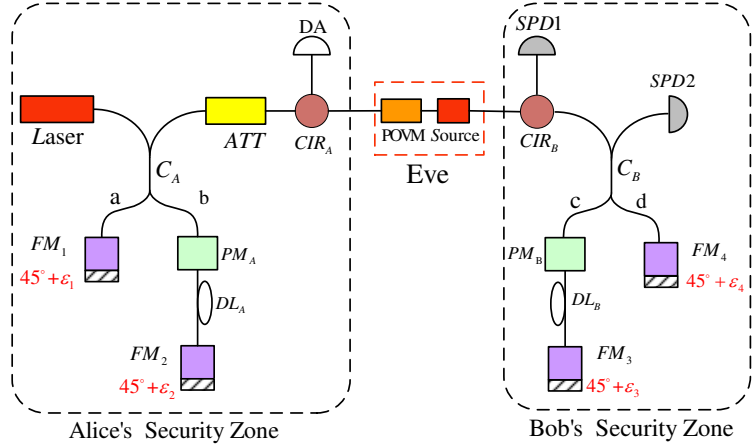}
 \caption{\label{fig1}The simple diagram of the FMQC system and Eve's attack. $\varepsilon_1$, $\varepsilon_2$, $\varepsilon_3$, $\varepsilon_4$ are rotation angle errors of $FM_1$, $FM_2$, $FM_3$, $FM_4$, respectively. Laser, pulse laser diode; ATT, attenuator; DL, delay line; CIR, circulator; DA, Alice's detector to monitor possible Trojan-horse photons; Source, Eve's light source to send photons to Bob; SPD, single-photon detector; other abbreviations defined in text.}
 \end{figure}

In this paper, we study the security of FMQC system against
imperfect Faraday mirrors. We calculate the changed signal states
caused by two imperfect FMs in Alice's security zone. Then a passive
Faraday-mirror(PFM) attack is implemented in this system. By
changing the values of the imperfection parameters of Alice's FMs,
we calculate the quantum bit error rate induced by Eve and the
probability that Eve obtains outcomes successfully. We can conclude
that the imperfection of the FM on the path where there is a phase
modulator makes the system sensitive to the attack. Since PFM attack
is just an individual attack based on intercept-resend attack, it is
not optimal.  Thus at last we do a security analysis taking the
imperfections of FMs in both Alice's and Bob's security zones into
account. A modified key rate as a function of the Faraday mirror
imperfections is given and the analysis indicates that both Alice's
and Bob's imperfect FMs can compromise the secure key.

This paper is organized as follows: In Sec. \uppercase\expandafter{\romannumeral2} we introduce the imperfection
of FMs and the resulting loophole. In Sec. \uppercase\expandafter{\romannumeral3}  we simulate a PFM attack which
can distinguish four states sent by Alice based on her imperfect FMs. In Sec. \uppercase\expandafter{\romannumeral4}
we analyze the security of the system against both Alice's and Bob's imperfect FMs and simulate the key rate taking the Faraday mirror imperfections as parameters. Finally, a brief conclusion of this paper is present in Sec. \uppercase\expandafter{\romannumeral5}.

\section{security loopholes induced by imperfect Faraday-mirrors}

In above discussion, the angle of Faraday rotator is thought to be exactly $45^{\circ}$. But in fact, the angle
always has an error $\varepsilon$ and the Jones matrix of a practical FM is given by
 $FM(\varepsilon)
= -\left(
\begin{array}{ccc}
 \sin(2\varepsilon) & \cos(2\varepsilon)  \\
 \cos(2\varepsilon) & -\sin(2\varepsilon)
\end{array} \right)$.
 Generally speaking, the maximal rotation angle error tolerance is $1^{\circ}$. When FMs are imperfect, the
birefringence effect of fibers cannot be compensated totally and additional QBER will be induced. What's worse,
the imperfection of Alice's FMs will leave a loophole for Eve to obtain more information about the secure key.

For multi-photon pulses, Eve can take a photon-number-splitting
(PNS) attack \cite{16,17,18} where she maintains one photon to
measure and lets other photons pass through. Here we only consider
the single-photon case. Simply and without losing generality, assume
that the incoming state is the horizontal polarization state, i.e,
$|\psi_{in}\rangle = [1~~0]^{T}$. Then the Jones vectors of the
output polarization states for the two time modes are given by
\begin{equation}
\begin{aligned}
|\psi_{a}\rangle&=-\left(\begin{array}{ccc}
 \sin(2\varepsilon_1) & \cos(2\varepsilon_1)  \\
 \cos(2\varepsilon_1) & -\sin(2\varepsilon_1)
\end{array} \right) |\psi_{in}\rangle=
-\left(
\begin{array}{ccc} \sin(2\varepsilon_{1})  \\
\cos(2\varepsilon_{1})
\end{array} \right),\\
|\psi_{b}\rangle&=-\left(
\begin{array}{ccc}
 e^{ik\delta_a} & 0  \\
 0 & 1
\end{array} \right)
\left(
\begin{array}{ccc}
 \sin(2\varepsilon_2) & \cos(2\varepsilon_2)  \\
 \cos(2\varepsilon_2) & -\sin(2\varepsilon_2)
\end{array} \right) \left(\begin{array}{ccc}
 e^{ik\delta_a} & 0  \\
 0 & 1
\end{array} \right)|\psi_{in}\rangle\\
&=-\left(
\begin{array}{ccc} \sin(2\varepsilon_{2})e^{2ik\delta_a}  \\
\cos(2\varepsilon_{2})e^{ik\delta_a}
\end{array} \right),
\end{aligned}
\end{equation}
where $k=0,1,2,3$, and $|\psi_{a}\rangle$, $|\psi_{b}\rangle$ are the Jones vectors of time mode $a$ and $b$,
respectively. $\delta_a$ is the phase modulated by the phase modulator and $\varepsilon_{1}$, $\varepsilon_{2}$ are the rotation angle errors of $FM_{1}$ and $FM_{2}$, respectively. When the FMs
are imperfect, the states sent by Alice are not the standard BB84 states, $|\psi_{k}\rangle= (|a\rangle+e^{ik\delta_a}|b\rangle) / \sqrt{2}$, where $\delta_a=\pi/2$. The four new states are given by \begin{eqnarray}
|\Phi_{k}\rangle&=&[\sin(2\varepsilon_{1})|Ha\rangle+\cos(2\varepsilon_{1})|Va\rangle+\sin(2\varepsilon_{2}) e^{i2k\delta_a}
|Hb\rangle+\cos(2\varepsilon_{2})e^{ik\delta_a}|Vb\rangle
]/\sqrt{2}.
\end{eqnarray} We can see that the four new states are in three-dimensional Hilbert space. To show it more
clearly, we let $|H \rangle =cos(2\varepsilon_{1})| X\rangle+sin(2\varepsilon_{1})| Y\rangle$ and
$| V \rangle=-sin(2\varepsilon_{1})| X\rangle+cos(2\varepsilon_{1})| Y\rangle$, and denote $|Xb\rangle=|x_{1}\rangle$,
$|Yb\rangle=|x_{2}\rangle$, $|Ya\rangle=|x_{3}\rangle$. Then the four new states can be rewritten as
\begin{eqnarray}
|\Phi_{k}\rangle&=&\{[\sin(2\varepsilon_{2})\cos(2\varepsilon_{1})e^{i2k\delta_a}-\sin(2\varepsilon_{1})\cos(2\varepsilon_{2})e^{ik\delta_a}]|x_{1}\rangle\nonumber\\
&&+[\sin(2\varepsilon_{2})\sin(2\varepsilon_{1})e^{i2k\delta_a}+\cos(2\varepsilon_{2})\cos(2\varepsilon_{1})e^{ik\delta_a}]|x_{2}\rangle+|x_{3}\rangle\}/\sqrt{2},\end{eqnarray}
which is quite different from Eq. (9) of Ref. [13]. The dimension of
Hilbert space of the states sent by Alice is 3 instead of 2, which
will give Eve more information about the secure key.

By calculating the inner products between any two of the four states
we can describe them in three-dimensional
 Hilbert space as shown in Fig. 2. The inner products between any two of the four states are given by
\begin{equation}
\begin{aligned}
\langle\Phi_{k}|\Phi_{k+1}\rangle&=\frac{1}{2}[\cos^{2}(2\varepsilon_{2})-i\cos^{2}(2\varepsilon_{2})],~k=0,1,2,3,\\
\langle\Phi_{0}|\Phi_{2}\rangle&=\langle\Phi_{1}|\Phi_{3}\rangle=\sin^{2}(2\varepsilon_{2}).
\end{aligned}
\end{equation}

\begin{figure}[!t]
 \includegraphics*[scale=0.82]{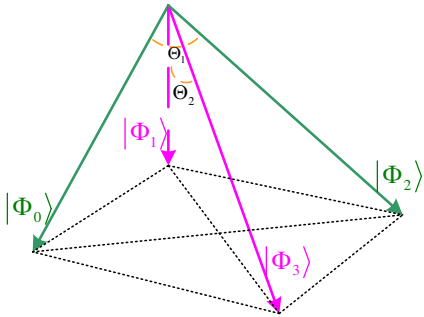}
 \caption{\label{fig3}The distribution of the four states in three-dimensional space. $\Theta_{1}$ is the angle between $|\Phi_{0}\rangle$ and $|\Phi_{2}\rangle$ and
 $\Theta_{2}$ is the angle between $|\Phi_{1}\rangle$ and $|\Phi_{3}\rangle$. $|\Phi_{0}\rangle$ and $|\Phi_{2}\rangle$ are in the bisector plane of the angle between $|\Phi_{1}\rangle$ and
 $|\Phi_{3}\rangle$.
 In the same way, $|\Phi_{1}\rangle$ and $|\Phi_{3}\rangle$ are in the bisector plane of the angle between $|\Phi_{2}\rangle$ and $|\Phi_{4}\rangle$.}
 \end{figure}
We find that all the inner products only relate to $\varepsilon_{2}$
which means $\varepsilon_{1}$ doesn't
 compromise the security of the system. This is easy to understand because $\varepsilon_{2}$ is the angle error of $FM_{2}$ on the path where the pulse is encoded and $\varepsilon_{1}$ is the angle error of $FM_{1}$ on the path where the pulse is just reference pulse. We can also conclude that only $\varepsilon_{3}$ will open a loophole in Bob's security zone from the symmetry between Alice and Bob.

\section{Passive Faraday-mirror attack based on Alice's imperfect Faraday-mirrors}
\subsection{Passive Faraday-mirror attack based on an intercept-resend attack}
Since the states sent by Alice are in three-dimensional Hilbert
space due to her imperfect FMs, Eve can use the operators belonging
to three-dimensional Hilbert space to measure them. Eve can make the
following attack: she intercepts each pulse from Alice's zone and
measures it with five POVM operators $\{F_{vac},F_{0},F_{1},
F_{2},F_{3}\}$ which satisfy the condition that $F_{vac} +
\sum_{k=0}^3 F_{k}= I$, where $I$ is identity matrix. When Eve
obtains the outcome corresponding to $F_{k}$ she resends a standard
BB84 state $|\psi_{k}\rangle= (|a\rangle+e^{ik\delta_a}|b\rangle) /
\sqrt{2}$. Here the POVM operator $F_{vac}$ corresponds to a vacuum
state.

In general, the main object of Eve is to find a set of POVM
operators that can minimize the quantum bit error rate between Alice
and Bob induced by her attack \cite{10}. Thus Eve can use this
specific strategy to minimize the QBER:  let
$\rho_{k}=|\Phi_{k}\rangle\langle\Phi_{k}|$,
$\rho=\sum_{k=0}^{3}\rho_{k}$,
 $L_{k}=\frac{1}{2}\rho_{k+1}+\rho_{k+2}+\frac{1}{2}\rho_{k+3}$,
$F_{k}=r\rho^{-1/2}|E_{k}\rangle\langle E_{k}|\rho^{-1/2}$, where $|E_{k}\rangle$ is the eigenvector of matrix
$\rho^{-1/2} L_{k} \rho^{-1/2}$ corresponding to the minimal nonzero eigenvalue and $r$
is the maximal real number ensuring that the matrix $F_{vac} =I-\sum_{k=0}^3 F_{k}$ is positive. Here, we use
five POVM operators to distinguish all the four states sent by Alice instead of only distinguishing two states
$\{\Phi_0, \Phi_3\}$ in Ref. [13]. The PFM attack proposed in Ref. [13] is combined with the phase-remapping
attack which can make $\delta_a\in[0,\pi/2]$ and when $\delta_a\neq\pi/2$ only $F_0$ and $F_3$ can be used to
minimize the QBER. But in our attack $\delta_a$ can only be $\pi/2$, thus we can use $F_0$, $F_1$, $F_2$ and
$F_3$ to distinguish all the four states sent by Alice and minimize the QBER simultaneously.

We only focus on two main parameters: the quantum bit error rate, $QBER$, between Alice and Bob induced by Eve's attack, and the probability that Eve obtains an outcome corresponding to $F_{k}$ successfully, $P_{succ}$. They are defined as follow:
 \begin{equation}
 QBER=\frac{\sum_{k=0}^3 Tr(F_{k}L_{k})}{\sum_{k=0}^3
 Tr(F_{k}\rho)},
 \end{equation}
 \begin{equation}
 P_{succ}=\frac{1}{4}\sum_{k=0}^3 Tr(F_{k}\rho).
 \end{equation}

\subsection{Simulation}
Since the system is one-way, $\delta_a=\pi/2$. As the maximal
rotation angle error tolerance is $1^{\circ}$, we let
$\varepsilon_{1}$ and $\varepsilon_{2}$ both change from
$-1^{\circ}$ to $1^{\circ}$. The results are shown in Fig. 3. From
Fig. 3~(a), we can see that the QBER is almost constant no matter
how $\varepsilon_{1}$ or $\varepsilon_{2}$ changes. In fact, the
QBER changes with $\varepsilon_{1}$ or $\varepsilon_{2}$ slightly
 and the difference is so small that we can ignore it. As for $P_{succ}$, Fig. 3~(b) shows how $P_{succ}$
 changes with $\varepsilon_{1}$ and $\varepsilon_{2}$. We can conclude that $P_{succ}$ depends on
$\varepsilon_{2}$ alone. From Fig. 3~(c) we can see that when
$\varepsilon_{2}$ is given, $P_{succ}$ is almost constant however
$\varepsilon_{1}$ changes and Fig. 3~(d) shows that the bigger the
absolute value of $\varepsilon_{2}$ is (which means the angle error
of $FM_{2}$ is bigger), the bigger the $P_{succ}$ is. This can be
easily explained by the results of inner products between Alice's
four states in Sec \uppercase\expandafter{\romannumeral2}. When
$|\varepsilon_{2}|$ gets bigger, each
$\langle\Phi_{k}|\Phi_{k+1}\rangle$ gets smaller which means the
differences between the four states are bigger. Therefore,
$P_{succ}$ gets bigger as  $|\varepsilon_{2}|$ gets bigger.
 \begin{figure}[!t]
 \centering
 \includegraphics*[scale=0.32]{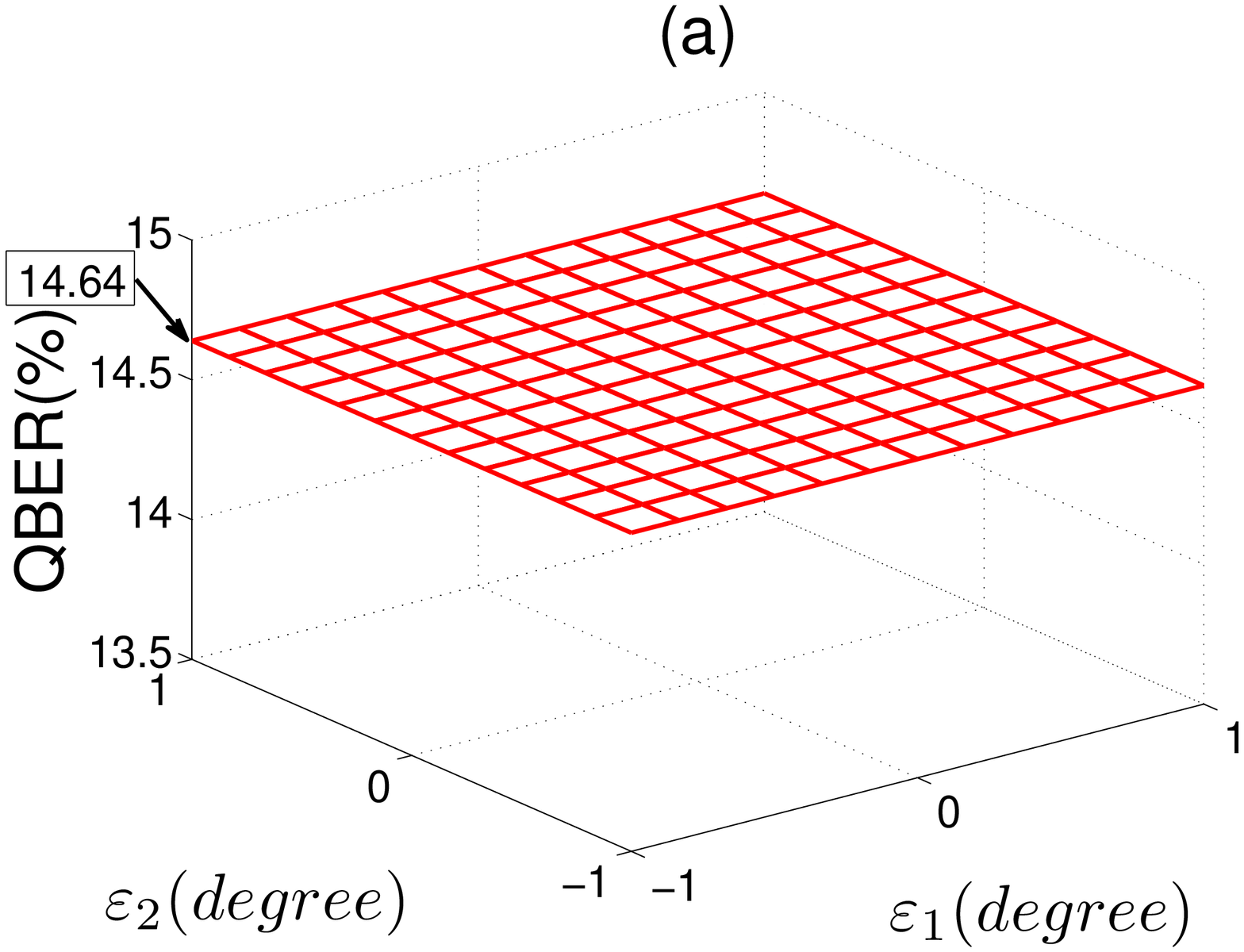}
 \includegraphics*[scale=0.32]{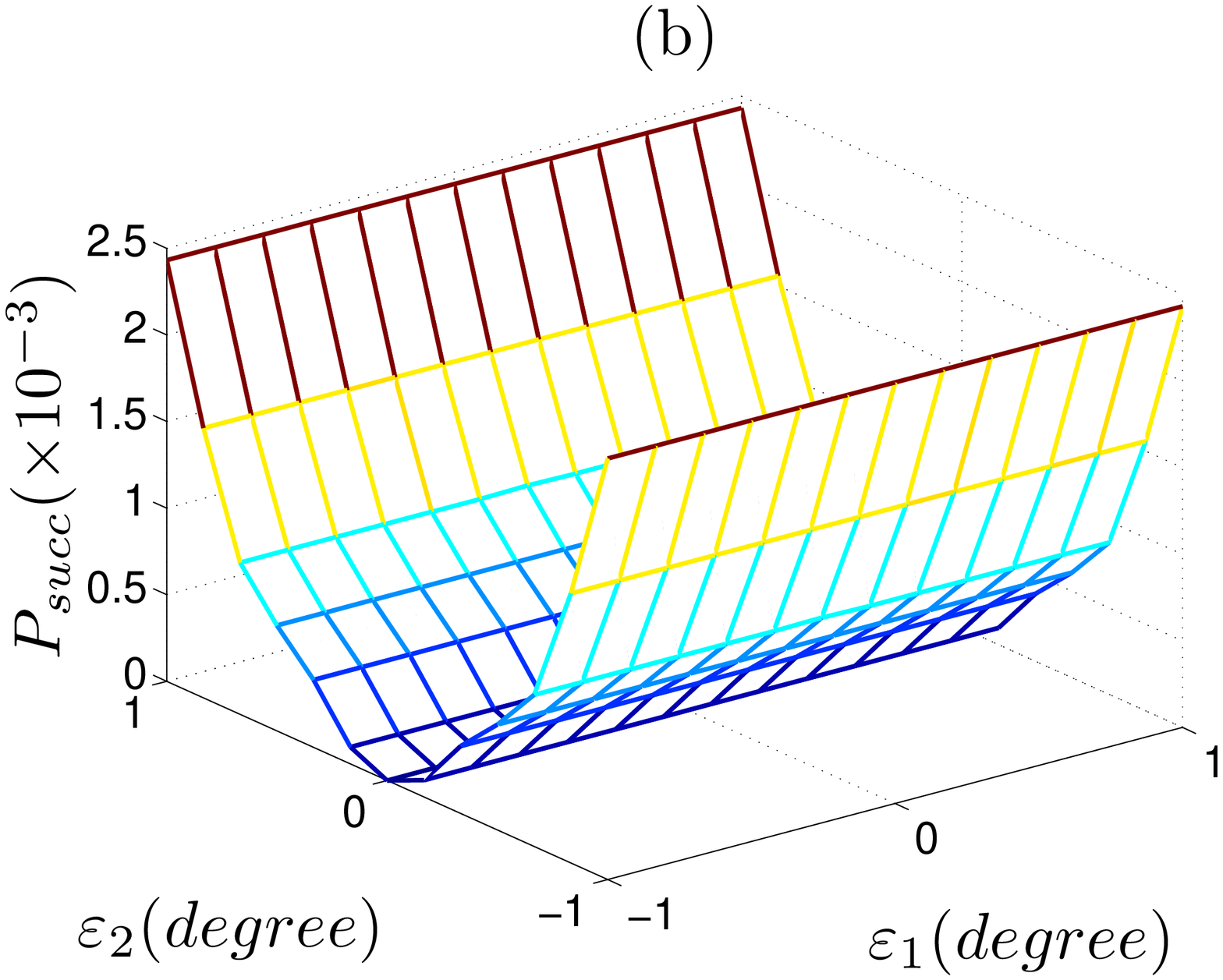}
 \includegraphics*[scale=0.32]{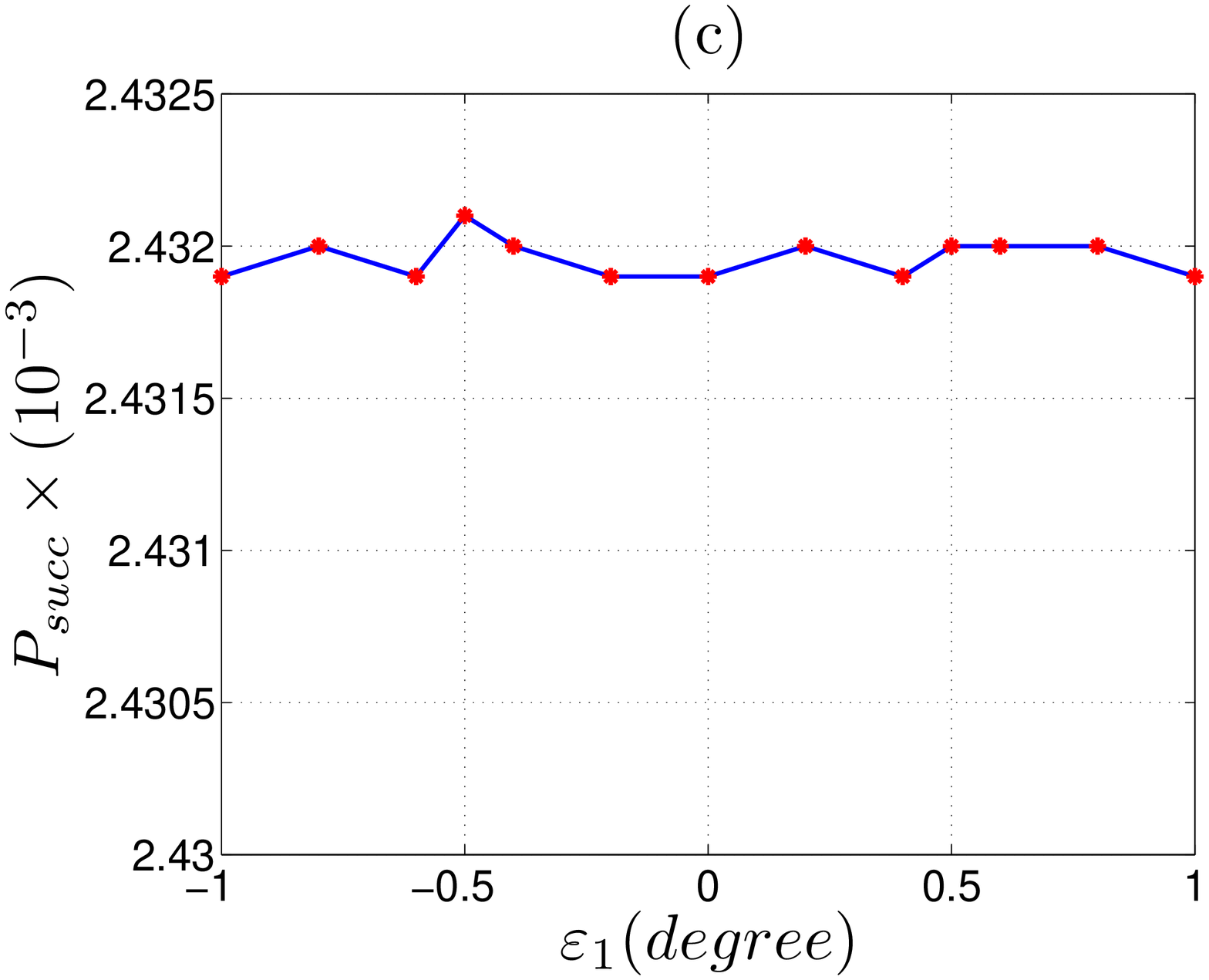}
 \includegraphics*[scale=0.32]{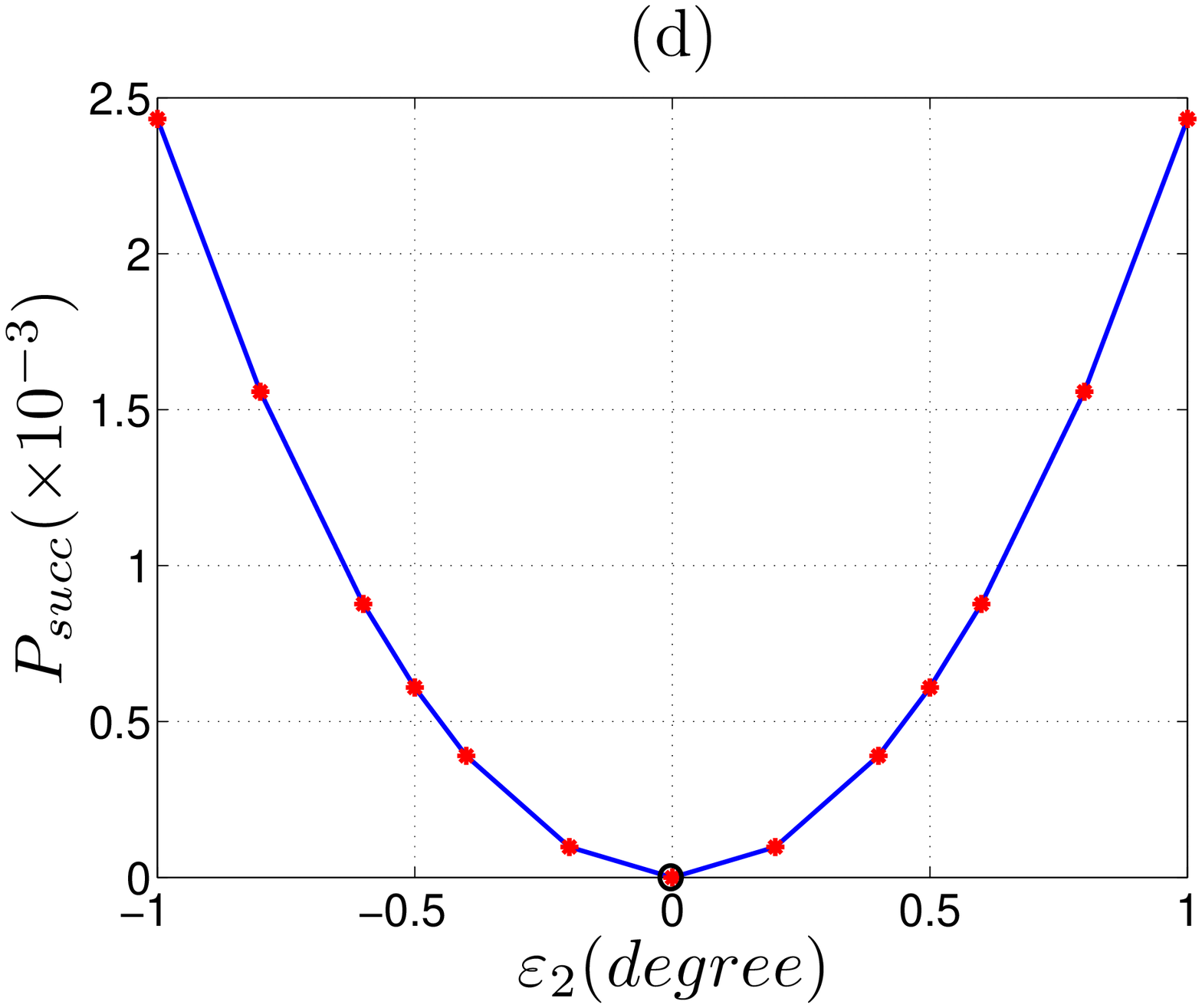}
 \caption{\label{fig2} (a) The relationship between QBER and $\varepsilon_{1},
 \varepsilon_{2}$. Note that, QBER is almost constant. (b) The relationship between $P_{succ}$ and $\varepsilon_{1}, \varepsilon_{2}$.
(c) The relationship between $P_{succ}$ and $\varepsilon_{1}$ when
$\varepsilon_{2}=1^{\circ}$. Note that, when $\varepsilon_{2}$ is
given, $P_{succ}$ is almost constant and changes very slightly with
$\varepsilon_{1}$. (d) The relationship between $P_{succ}$ and
$\varepsilon_{2}$ when $\varepsilon_{1}=1^{\circ}$. Note that, the
bigger the absolute value of $\varepsilon_{2}$ is, the bigger
$P_{succ}$ is. The special point that $\varepsilon_{2}=0^{\circ}$ is
not considered in our simulation.}
 \end{figure}

Moreover, $\varepsilon_{2}$ can't be $0^{\circ}$ because if $\varepsilon_{2}=0^{\circ}$, the density operator
$\rho$ becomes
\begin{equation}
\rho=2\left(
\begin{array}{ccc}
 \sin^{2}(2\varepsilon_{1}) & -\sin(2\varepsilon_{1})\cos(2\varepsilon_{1})& 0  \\
 -\sin(2\varepsilon_{1})\cos(2\varepsilon_{1})&
 \cos^{2}(2\varepsilon_{1})&0 \\0&0&1
\end{array} \right).
\end{equation}
Then we can find that its rank is 2 which means the dimension of the
Hilbert space becomes 2 instead of 3, thus the PFM attack is not
effective.

\section{Security analysis against imperfections of Faraday-mirrors}
\subsection{Security analysis against four imperfect Faraday-mirrors}
In this FMQC system, there are two FMs in Alice' and Bob's zones,
respectively. If Eve is the manufacturer of Alice's and Bob's
instruments, she can set the rotation angle errors of the four FMs
before providing the instruments to them. Thus we want to know
whether Eve can take use of these four imperfect FMs to obtain more
information about the secure key without the legitimate users'
awareness.

Firstly, assume that all the FMs in the system are perfect and Alice
sends the horizontal polarization state. When there' s no Eve and
the quantum channel is noiseless, the pulses that pass through
$(S_a, L_b)$ and $(L_a, S_b)$ are
$|\psi_{S_a,L_b}\rangle=e^{ik\delta_b}\left(
\begin{array}{ccc}
 1  \\
 0
\end{array} \right)$ and
$|\psi_{L_a,S_b}\rangle=e^{ik\delta_a}\left(
\begin{array}{ccc}
 1  \\
 0
\end{array} \right)$ when they arrive at $C_B$, respectively. $S_a$, $L_a$, $S_b$, $L_b$ are denoted as the
short and long arm of Alice's and Bob's zone, respectively.

Then if the four FMs are imperfect and Eve knows their rotation
angle errors are $\varepsilon_1$, $\varepsilon_2$, $\varepsilon_3$
and $\varepsilon_4$, respectively. Also when Alice sends the
horizontal polarization state and the quantum channel is noiseless,
the pulses that pass through $(S_a, L_b)$ and $(L_a, S_b)$ become
\begin{equation}
\begin{aligned}
&|\psi_{S_a,L_b}^{'}\rangle=\left(\begin{array}{ccc}
 e^{i2k\delta_b}sin(2\varepsilon_1)sin(2\varepsilon_3)+e^{ik\delta_b}cos(2\varepsilon_1)cos(2\varepsilon_3)  \\
 e^{ik\delta_b}sin(2\varepsilon_1)cos(2\varepsilon_3)-cos(2\varepsilon_1)sin(2\varepsilon_3)
\end{array} \right),\\
&|\psi_{L_a,S_b}^{'}\rangle=\left(\begin{array}{ccc}
 e^{i2k\delta_a}sin(2\varepsilon_2)sin(2\varepsilon_4)+e^{ik\delta_a}cos(2\varepsilon_2)cos(2\varepsilon_4)  \\
 e^{i2k\delta_a}sin(2\varepsilon_2)cos(2\varepsilon_4)-e^{ik\delta_a}cos(2\varepsilon_2)sin(2\varepsilon_4)
\end{array} \right),
\end{aligned}
\end{equation}when they arrive at $C_B$, respectively.
If Eve doesn't want Alice and Bob to know the change of the states,
she must control the rotation angle errors of the four FMs to always
satisfy
\begin{equation}
\begin{aligned}
&|\psi_{S_a,L_b}^{'}\rangle=|\psi_{S_a,L_b}\rangle,\\
&|\psi_{L_a,S_b}^{'}\rangle=|\psi_{L_a,S_b}\rangle.
\end{aligned}
\end{equation}
Then we can find a solution to Eq. (11),
\begin{equation}
\begin{aligned}
sin(2\varepsilon_1)=e^{-ik\delta_b}sin(2\varepsilon_3),~cos(2\varepsilon_1)=cos(2\varepsilon_3),\\
sin(2\varepsilon_2)=e^{-ik\delta_a}sin(2\varepsilon_4),~cos(2\varepsilon_2)=cos(2\varepsilon_4).
\end{aligned}
\end{equation}

Since the value of $k$ varies from \{0,1,2,3\} constantly and
randomly when the system is running, Eve cannot set $\varepsilon_1$,
$\varepsilon_2$, $\varepsilon_3$, $\varepsilon_4$ to always satisfy
Eq. (12). Alice and Bob must be able to perceive the change of
states. Thus Eve can't take use of four imperfect FMs to obtain more
information about the secure key without the Alice's and Bob's
awareness.

\subsection{Security analysis of phase-encoded BB84 protocol against
imperfect Faraday-mirrors} The PFM attack in the previous section is
an individual attack based on intercept-resend attack, thus it is
not necessarily the optimal attack. Besides, since the minimal QBER
induced is about $14.64\%$ which is just the tolerable upper bound
of error rate in individual attack, it is not realizable in this
system. What legal parties want is to make their protocol secure
against any attack permitted by quantum mechanics. Thus we need to
perform a compact security analysis.

In FMQC system, to implement a phase-encoded BB84 protocol, Alice
encodes the signals in the $X$ basis when she chooses the phase
$\varphi_A\in\{0, \pi\}$ and in the $Y$ basis when she chooses
$\varphi_A\in\{\pi/2, 3\pi/2\}$. Bob decodes the signals in the same
way as Alice does. In our scheme, the error rate $\delta_X$ and the
fraction $q_X$ of nonvacuum events are estimated when both Alice and
Bob choose the $X$ basis. The error rate $\delta_{ph}$ and fraction
$q_Y$ of nonvacuum events are estimated when both Alice and Bob
choose the $Y$ basis. The error rate $\delta_Y$ and fraction
$q_{ph}$ are estimated when Alice chooses the $X$ basis and Bob
chooses the $Y$ basis. The final secure key is only extracted from
the data measured when Alice and Bob both choose the $X$ basis. When
Alice's and Bob's FMs are imperfect, the states prepared by Alice
will be different from the standard BB84 states as shown in the
above section and the bases chosen by Bob will also be different
from the perfect $X$, $Y$ basis. Thus the secure key rate will be
compromised.

Here we follow the security proof proposed in Ref. [19]. We assume
that the channel is symmetric which means $q_X=q_Y=q_{ph}=q$ and
ignore the imperfections of detectors. Then the key rate in an
infinite length limit is given by
\begin{equation}
\begin{aligned}
R_X\geq 1-h({\delta_{ph}})-h(\delta_X),
\end{aligned}
\end{equation}
where $\delta_{ph}=min\{\frac{1}{2},~\delta_Y+8\frac{\Delta}{q}[(1-\frac{\Delta}{q})(1-2\delta_Y)+\sqrt{\frac{\Delta}
{q}(1-\frac{\Delta}{q})\delta_Y(1-\delta_Y)}]\}$, and  $\Delta=\frac{1}{2}[1-F(\rho_X,\rho_Y)]$. $F(\rho_X,\rho_Y)\equiv Tr(\sqrt{\rho_X}\rho_Y\sqrt{\rho_X})^{\frac{1}{2}}$ is the fidelity which characterizes the basis dependence of the source.

Since we can conclude that $\varepsilon_1$ and $\varepsilon_4$ will
not compromise the security of the system, we let
$\varepsilon_1=\varepsilon_4=0$ when simulating the key rate. In Eq.
(13), $\Delta$ is related to $\varepsilon_2$ and nonzero
$\varepsilon_3$ can increase $\delta_X$ and $\delta_Y$. Thus the
imperfections of Alice's and Bob's FMs have influences on the final
key rate.

We define $\rho_X=\frac{1}{2}(|\Phi_{0}\rangle\langle\Phi_{0}|+|\Phi_{2}\rangle\langle\Phi_{2}|)$, $\rho_Y=\frac{1}{2}(|\Phi_{1}\rangle\langle\Phi_{1}|+|\Phi_{3}\rangle\langle\Phi_{3}|)$, $\delta_X=\delta+\delta_{X,\varepsilon_3}$, $\delta_Y=\delta+\delta_{Y,\varepsilon_3}$, where $\delta$ is the inherent error rate and $\delta_{X,\varepsilon_3}$, $\delta_{Y,\varepsilon_3}$ are the error rates induced by $\varepsilon_3$ in the $X$ and $Y$ bases, respectively. When $\varepsilon_3\neq 0$, the bases used by Bob are not standard $X$, $Y$ basis and they are given by
\begin{equation}
|\Phi^{'}_{k}\rangle=\frac{1}{\sqrt{2}}
\left(\begin{array}{c}\sin(2\varepsilon_3)\\e^{\frac{ik\pi}{2}}\cos(2\varepsilon_3)\\1
\end{array} \right),
\end{equation}
where $k=0,1,2,3$. Then the error rates induced by $\varepsilon_3$ are given by
\begin{equation}
\begin{aligned}
\delta_{X,\varepsilon_3}=\frac{1}{2}(\langle\Phi_{2}|\Phi^{'}_{0}\rangle\langle\Phi^{'}_{0}|\Phi_{2}\rangle+
\langle\Phi_{0}|\Phi^{'}_{2}\rangle\langle\Phi^{'}_{2}|\Phi_{0}\rangle),\\
\delta_{Y,\varepsilon_3}=\frac{1}{2}(\langle\Phi_{3}|\Phi^{'}_{1}\rangle\langle\Phi^{'}_{1}|\Phi_{3}\rangle+
\langle\Phi_{1}|\Phi^{'}_{3}\rangle\langle\Phi^{'}_{3}|\Phi_{1}\rangle).
\end{aligned}
\end{equation}

\begin{figure}[!t]
 \includegraphics*[scale=0.35]{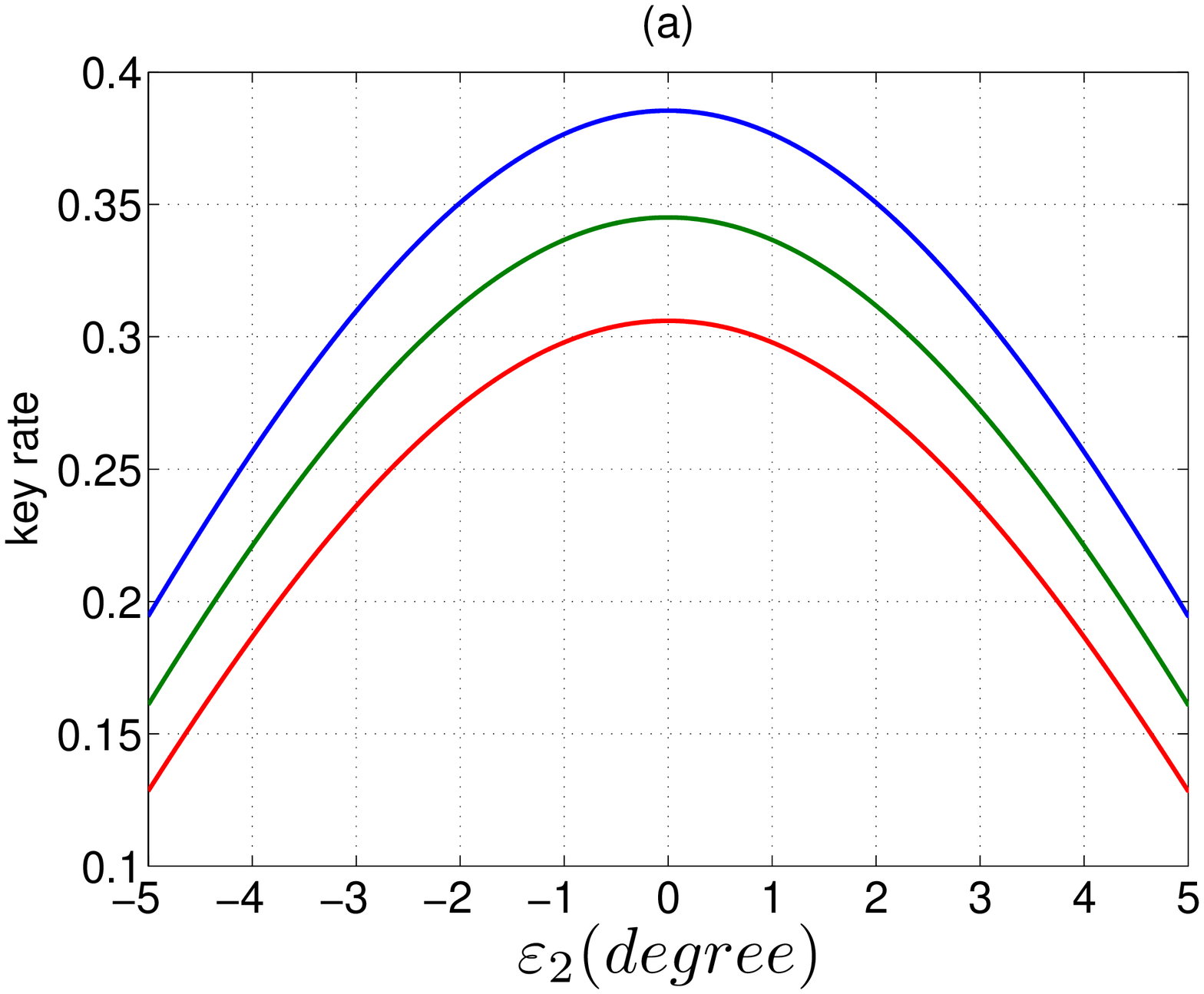}
 \includegraphics*[scale=0.35]{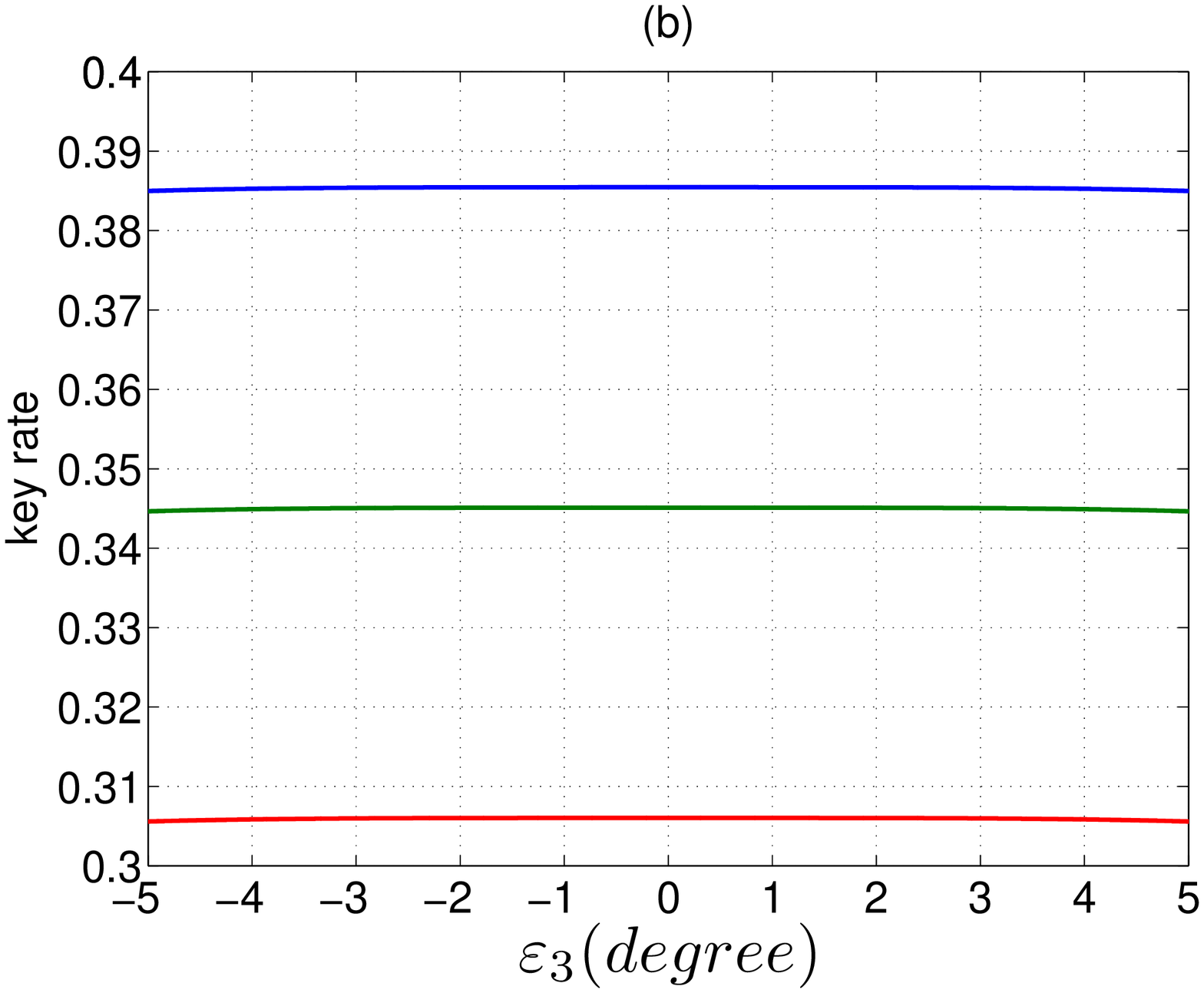}
 \caption{\label{fig2} In both plots, the blue, green, red lines are plotted with $q=0.5$ and
 $\delta=5.5\%, 6\%, 6.5\%$, respectively. (a) The relationship between key rate and $\varepsilon_{2}$.
 We ignore the influence of $\varepsilon_3$ which means $\delta_X=\delta_Y=\delta$. (b) The relationship between key rate and $\varepsilon_{3}$. We ignore the influence of $\varepsilon_2$ ($\varepsilon_2=0$).}
 \end{figure}
Fig. 4~(a) and (b) show how $\varepsilon_2$ and $\varepsilon_3$
compromise the final secure key rate, respectively. To show the
effect of $\varepsilon_2$ and $\varepsilon_3$ more clearly, we let
both of them change from $-5^\circ$ to $5^\circ$. From Fig. 4~(a) we
can see that the key rate varies significantly as $\varepsilon_2$
changes and the effect on key rate gets more significant as $\delta$
gets smaller. Fig. 4~(b) shows that the key rate varies very
slightly as $\varepsilon_3$ changes. We can conclude that the key
rate decreases as the absolute values of $\varepsilon_2$,
$\varepsilon_3$ get bigger and $\varepsilon_2$ has a much more
remarkable effect on key rate than $\varepsilon_3$.

\section{CONCLUSION}
The one-way Faraday-Michelson quantum cryptography system is a very
useful practical system where FMs are used to compensate for the
birefringence of fiber. However, practical FMs are always imperfect
which means the angle of Faraday rotator is not exactly $45^\circ$.
We analyze the security of the system taking Alice's and Bob's
imperfect Faraday mirrors into account. Passive Faraday-mirror
attack is proposed in two-way system. We consider this attack in
this one-way FMQC system. By changing the values of the imperfection
parameters of Alice's FMs, we calculate the quantum bit error rate
between Alice and Bob induced by Eve's attack and the probability
that Eve obtains outcomes successfully. Using simulation we find
that only the imperfect FM on the path where the encoded signal
pulse is transmitted can give Eve more information of the secure key
and the other FM in Alice's security zone has nothing to do with the
attack. In our attack, Eve can use five POVM operators belonging to
three-dimensional Hilbert space to distinguish all the four states
sent by Alice. Since PFM attack is just an individual attack based
on intercept-resend attack and the minimal QBER is about $14.64\%$
which is just the tolerable upper bound of error rate in individual
attack, it is neither optimal nor realizable in this QKD system.
Then a modified key rate as a function of the Faraday mirror
imperfections is given and it is secure against any attack permitted
by quantum mechanics. The security analysis indicates that both
Alice's and Bob's imperfect FMs can compromise the secure key and
Alice's imperfect FM has a much more remarkable effect on key rate
than Bob's. The imperfection of the FMs can remind the system's
manufacturer to use as good a FM as possible and the security
analysis can tell legal parties how to adjust privacy amplification
to keep a lookout for a potential eavesdropper.

\section*{Acknowledgements}
This work is supported by National Natural Science Foundation of China, Grant No.U1204602,
National High-Tech Program of China, Grant No.2011AA010803 and the Open Project Program of State Key Laboratory
of Mathematical Engineering and Advanced Computing, Grant No.2013A14.


\begin{thebibliography}{24}
   \bibitem[1]{1} C. H. Bennett and G. Brassard, Proceedings of the IEEE International Conference on Computers,
   Systems and Signal Processing, Bangalore, India (IEEE, New York, 1984), pp. 175¨C179.
   \bibitem[2]{2} A. K. Ekert, Phys. Rev. Lett. \textbf{67}, 661 (1991).
   \bibitem[3]{3} H. K. Lo and H. F. Chau, Science \textbf{283}, 2050 (1999).
   \bibitem[4]{4} P. W. Shor and J. Preskill, Phys. Rev. Lett. \textbf{85}, 441 (2000).
   \bibitem[5]{5} D. Mayers, J. Assoc. Comput. Mach. \textbf{48}, 351 (2001).
   \bibitem[6]{6} R. Renner, N. Gisin and B. Kraus, Phys. Rev. A \textbf{72}, 012332 (2005).
   \bibitem[7]{7} N. Gisin, G. Ribordy, W. Tittel and H. Zbinden, Rev. Mod. Phys. \textbf{74}, 145$-$190 (2002).
   \bibitem[8]{8} N. Gisin, S. Fasel, B. Kraus, H. Zbinden and G. Ribordy, Phys. Rev. A \textbf{73}, 022320 (2006).
   \bibitem[9]{9} V. Makarov and Dag R. Hjelme, J. Mod. Opt. \textbf{52}, 691 (2005).
   \bibitem[10]{10} C. H. F. Fung, B. Qi, K. Tamaki and H. K. Lo, Phys. Rev. A \textbf{75}, 032314 (2007).
   \bibitem[11]{11} V. Makarov, A. Anisimov and J. Skaar, Phys. Rev. A \textbf{74}, 022313 (2006).
   \bibitem[12]{12} L. Lydersen, C. Wiechers, C. Wittmann, D. Elser, J. Skaar and V. Makarov, Nat. Photonics \textbf{4}, 686 (2010).
   \bibitem[13]{13} Shi-Hai Sun, Mu-Sheng Jiang and Lin-Mei Liang, Phys. Rev. A \textbf{83}, 062331 (2011).
   \bibitem[14]{14} A. Muller, T. Herzog, B. Huttner, W. Tittel, H. Zbinden and N. Gisin, Appl. Phys. Lett. \textbf{70}, 793 (1997).

   \bibitem[15]{15} Xiao-Fan Mo, Bing Zhu, Zheng-Fu Han, You-Zhen Gui\ and Guang-Can Guo, Opt. Lett. \textbf{30}, 19 (2005).
   \bibitem[16]{16} B. Huttner, N. Imoto, N. Gisin and T. Mor, Phys. Rev. A \textbf{51}, 1863 (1995).
   \bibitem[17]{17} G. Brassard, N. L\"utkenhaus, T. Mor and B. C. Sanders, Phys. Rev. Lett. \textbf{85}, 1330 (2000).
   \bibitem[18]{18} N. L\"utkenhaus and M. Jahma, New J. Phys. \textbf{4}, 44 (2002).
   \bibitem[19]{19} $\varnothing$ystein Mar$\phi$y, L. Lydersen and J. Skaar, Phys. Rev. A \textbf{82}, 032337 (2010).
\end{thebibliography}
\end{document}